\documentstyle[twoside,fleqn,espcrc2,epsfig]{article}

\newcommand{\mH}{{\mathbf{H}}}
\newcommand{\dual}{\mbox{}^{\ast}}
\newcommand{\dd}{\mbox{\rm d}}
\newcommand{\dD}{{\cal D}}
\newcommand{\Z}{{Z \!\!\! Z}}
\newcommand{\beqn}{\begin{eqnarray}}
\newcommand{\eeqn}{\end{eqnarray}}
\newcommand{\eq}[1]{(\ref{#1})}
\newcommand{\cD}{{\cal D}}

\newcommand{\cZ}{{\cal Z}}

\title{
\thispagestyle{empty}
\vspace{-14mm}
\rightline{\small KANAZAWA-01-12}
\rightline{\small ITEP-LAT/2001-02}
\rightline{\small 15 October, 2001}
\vspace{2mm}
Numerical study of Frohlich and Marchetti monopole creation
operator\thanks{Presented by V.A.B. at Lattice 2001, Berlin.}}
\author{V.A.~Belavin\address{Institute of Theoretical and Experimental
Physics, B.Cheremushkinskaya 25, Moscow, 117259, Russia},
M.N.~Chernodub$^{\mathrm{a,}}$\address{Institute for Theoretical Physics,
Kanazawa University, Kanazawa 920-1192, Japan},
M.I.~Polikarpov$^{\mathrm{a}}$}

\begin{document}

\begin{abstract}
The monopole creation operator proposed recently by Frohlich and Marchetti
is investigated in the Abelian Higgs model with compact gauge field. We show
numerically that the creation operator detects the condensation of monopoles
in the presence of the dynamical matter field.
\end{abstract}

\maketitle

\section{INTRODUCTION}
The value of the deconfinement temperature is one of the most important
prediction of lattice QCD. To study the temperature phase transition we have
to investigate the order parameter. For full QCD when dynamical quarks are
taken into account, the string tension and the expectation value of the
Polyakov line are not the order parameters. If we accept the dual
superconductor model of QCD vacuum \cite{tHMa} we have the natural order
parameter for confinement -- deconfinement phase transition. This is the
value of the monopole condensate. It should be nonzero in the confinement
phase (the monopoles are condensed as Cooper pairs in ordinary
superconductor) and zero in the deconfinement phase. To extract monopole
from vacuum non-Abelian fields we have to perform the Abelian
projection~\cite{AbProj}, after that we can evaluate the value of the
monopole condensate using the monopole creation operator.

Originally the gauge invariant monopole creation operator was proposed by
Frohlich and Marchetti for compact $U(1)$ gauge theory~\cite{FrMa91}. The
construction is analogous to the Dirac creation
operator~\cite{DiracOperator} for a charged particle. The monopole
operator was numerically studied in compact Abelian gauge
model~\cite{PoPoWi91} as well as in the pure $SU(2)$ gauge theory both in
usual~\cite{ChPiVe95} and spatial~\cite{ChPoVe99} Maximal Abelian gauges. It
was found that the expectation value of this operator behaves as an order
parameter for confinement--deconfinement phase transition: the expectation
value is non-zero in the confinement phase and zero in the deconfinement
phase.  The similar conclusions were made for another types of the
monopole creation operators~\cite{DiGi}. These results confirm the dual
superconductor hypothesis~\cite{tHMa} for gluodynamics vacuum.

However, the monopole operator discussed in Ref.~\cite{FrMa91} exhibits some
inconsistency in the presence of charged matter fields, namely the Dirac
strings become visible. To get rid of the Dirac string dependence a new
monopole operator was proposed recently~\cite{FrMa99}. Note that even the
pure gluodynamics contains electrically charged fields in the Abelian
projection: the off--diagonal gluons are (doubly) charged with respect to
the diagonal gluon fields. Thus the newly proposed operator~\cite{FrMa99} is
more suitable for the investigation of confinement in $SU(N)$ gauge theories
then the older one~\cite{FrMa91}. The purpose of this paper is to check
whether the new monopole creation operator is the order parameter in
theories with matter fields. For simplicity we study the Abelian Higgs model
in the London limit having in mind the further numerical investigation of
the new monopole creation operator in non-Abelian gauge theories.

\vspace{-2mm}
\section{MONOPOLE OPERATORS}
The original version of the gauge invariant monopole creation operator
\cite{FrMa91} in compact $U(1)$ gauge theory is based on the duality of this
model to the Abelian Higgs model. The Higgs field $\phi$ is
associated with the monopole field and the non--compact dual gauge field
$B_\mu$ represents the dual photon. The gauge invariant operator which
creates the monopole in the point $x$, can be written as
the Dirac operator~\cite{DiracOperator} in the dual model:  \beqn
\Phi^{\mathrm{mon}}_x (H) = \phi_x \,
e^{i (B,H_x)}\,,
\label{Phi}
\eeqn
where the magnetic field of the monopole, $\mH$, is defined in the
three--dimensional time slice which includes the point $x$. By definition,
the magnetic monopole field satisfies the Maxwell equation, $\delta^{(3)}
H_x = \delta_x$ which guarantees the dual gauge invariance of the operator
$\Phi$ ($\delta^{(3)}$ is the three dimensional co-differential),
\beqn
\phi \rightarrow \phi e^{i \alpha}\,, \quad
B \rightarrow B + \dd \alpha.
\label{gauge:transformations}
\eeqn

The monopole creation operator \eq{Phi} can be re\-written in the original
representation in terms of the compact field $\theta$. In
lattice notations the expectation value of this operator is \cite{FrMa91}:
\beqn
\langle \Phi^{\mathrm{mon}} \rangle \! & = & \! {1 \over \cZ}
\int_{- \pi}^{\pi} \!\!\! \dD \theta \, \exp\{ -S(d \theta + W)\}\,,
\nonumber\\
\cZ \! & = & \! \int_{- \pi}^{ \pi} \!\!\! \cD \theta \,
\exp \{- S(d\theta) \}\,,
\label{original}
\eeqn
For compact lattice electrodynamics the general type of
the action satisfies the relation:  $S(d\theta+2\pi n)=S(d\theta)$, $n \in
\Z$. Besides the Coulomb monopole field $H$ the tensor form $W=2 \pi \delta
\Delta^{-1}(H_x - \omega_x)$ depends on the Dirac string $\omega$ which ends
at the monopole position, $\delta \dual \omega_x = \dual \delta_x$, and
is not restricted to the $3D$ time--slice.

The operator \eq{Phi} is well defined for the theories without dynamical
matter fields. However, if an electrically charged matter is added, then
the creation operator \eq{Phi} depends on the
positions of the Dirac strings. To see this fact we note that in the
presence of the dynamical matter the dual gauge field $B$ becomes compact.
Indeed, as we mentioned the pure compact gauge model is dual to the
non--compact $U(1)$ with matter fields (referred above as the (dual) Abelian
Higgs model). Reading this relation backwards we conclude that the presence of
the matter field leads to the compactification of the dual gauge field $B$.

The compactness of the dual gauge field implies automatically that the
gauge field transformation \eq{gauge:transformations} must be modified:
\beqn
\phi \rightarrow \phi e^{i \alpha}\,, \quad
B \rightarrow B + \dd \alpha + 2 \pi k\,,
\label{gauge:transformations:compact}
\eeqn
where the compactness of the gauge field, $B \in (-\pi,\pi]$, is supported
by the integer--valued vector field $k=k(B,\alpha)$.
The role of the field $k$ is to change shape of the dual Dirac strings
attached to the electric charges in the dual theory. One can easily check
that the operator \eq{Phi} is not invariant under the compact gauge
transformations \eq{gauge:transformations:compact}:  \beqn
\Phi^{\mathrm{mon}}_x (H) \to \Phi^{\mathrm{mon}}_x (H)
\, e^{2 \pi i (k,H_x)}\,.
\label{change}
\eeqn

This inconsistency is studied in Ref.~\cite{FrMa91}.  According to
eq.\eq{change} if the field $H$ is integer--valued then operator \eq{Phi} is
invariant under compact gauge transformations
\eq{gauge:transformations:compact}. This condition and the Maxwell equation
require for the field $H$ to have a form of a string attached to the
monopole ("Mandelstam string"): $H_x \to j_x$, $j \in \Z$. The string must
be defined in the three--dimensional time--slice similarly to the magnetic
field $H$. However, one can show that for a fixed string position the
operator $\Phi$ creates a state with an infinite energy. This difficulty may
be bypassed \cite{FrMa99} by summation  over all possible
positions of the Mandelstam strings with a suitable measure $\mu(j)$:
\beqn
\Phi^{\mathrm{mon,new}}_x = \phi_x \, \sum_{\stackrel{\dual j_x \in
\Z}{\delta \dual j_x = \delta_x}} \mu(j_x) \, e^{i (B,j_x)}\,.
\label{Phi:new}
\eeqn
If Higgs field $\phi$ is $q$--charged ($q \in \Z$), the
summation in eq.\eq{Phi:new} should be taken over $q$ different strings each
of which carries the magnetic flux $2 \pi \slash q$. The transformation of
$\Phi^{\mathrm{mon,new}}_x$ to the original representation can be easily
performed and we get the expression similar to eq.~\eq{original}.

\section{NUMERICAL RESULTS}

The purpose of the present publication is the numerical investigation of the
operator $\Phi^{\mathrm{mon,new}}_x$. We study the
monopole creation operator \eq{Phi:new} in the compact Abelian Higgs model
with the action:  \beqn S = - \beta \cos (\dd \theta) - \kappa \cos(\dd
\varphi + q \theta)\,, \eeqn where $\theta$ is the compact gauge field and
$\varphi$ is the phase of the Higgs field. For simplicity we consider the
London limit of the model in which the radial part of the Higgs field is
frozen. We calculate the (modified) effective
constraint potential,
\beqn V_{\mathrm{eff}}(\Phi) = - \ln \Bigl(\langle
\delta(\Phi - \Phi^{\mathrm{mon,new}}) \rangle\Bigr)\,.
\label{eff:potential}
\eeqn

\begin{figure}[!htb]
\vspace{2mm}
 \begin{minipage}{7.5cm}
 \begin{center}
  \epsfig{file=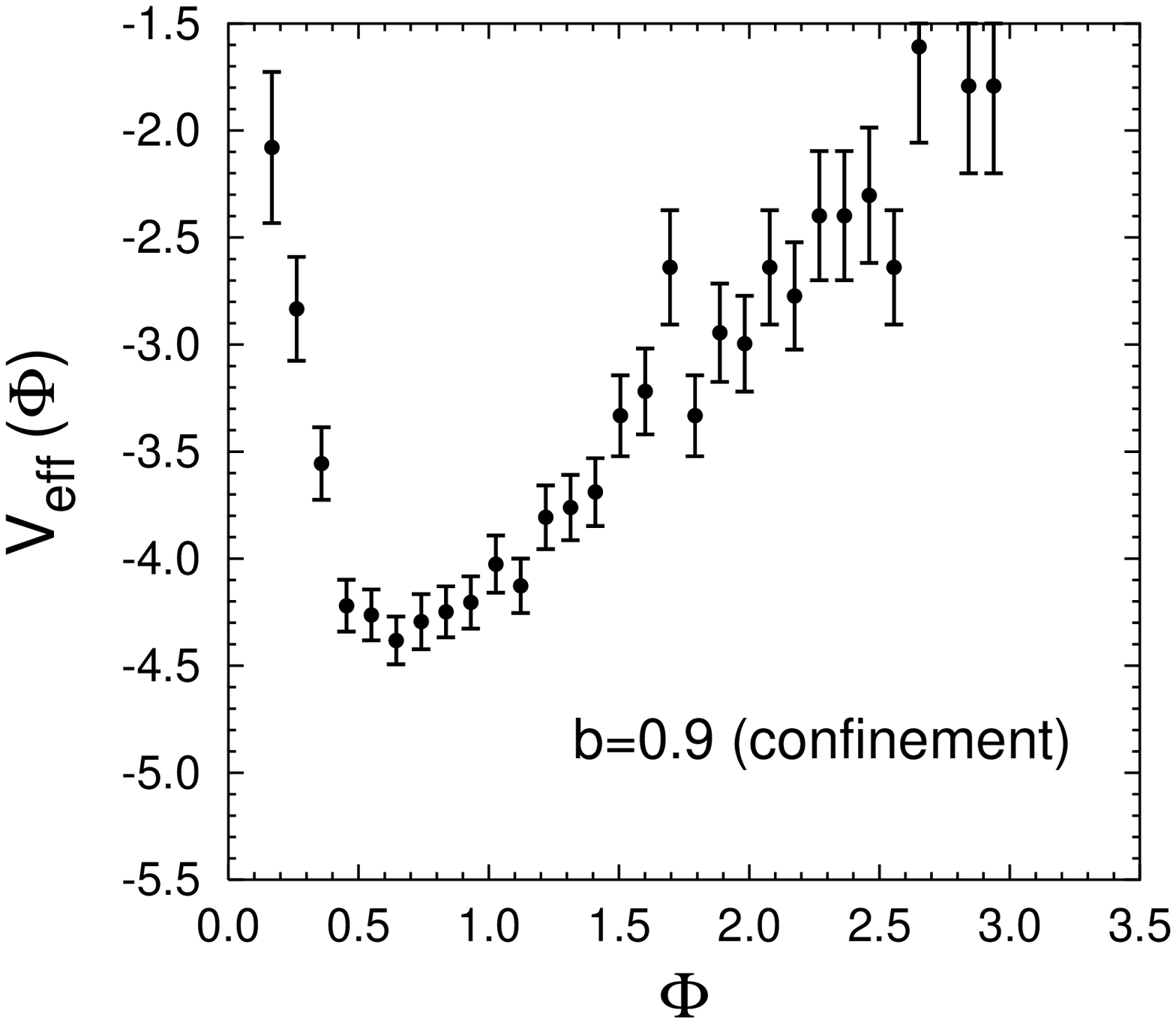,
  height=5.4cm}\\
(a) \\
\vskip 3mm
  \epsfig{file=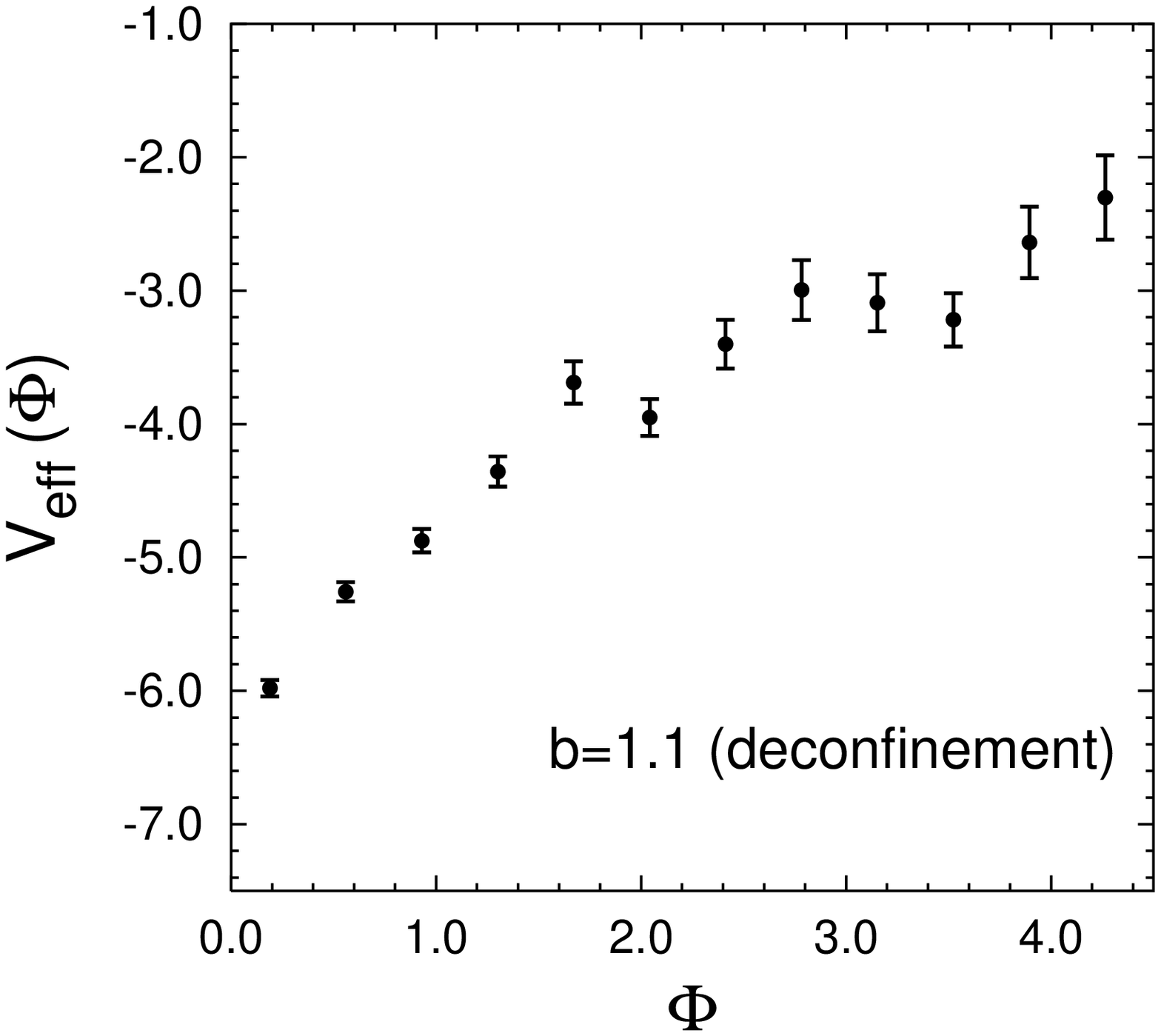,
  height=5.4cm}\\
(b)
 \end{center}
 \end{minipage}
\vspace{-8mm}
\caption{The effective monopole potential~\eq{eff:potential}
in (a) confinement and (b) deconfinement phases.}
\label{fig:potentials}
\vspace{-8mm}
\end{figure}
We simulate the theory on the $6^4$ lattice with the fixed $\kappa = 0.3$.
The larger charge, $q$, of the Higgs field, the easier the numerical
calculation of $V_{\mathrm{eff}}(\Phi)$. We performed our calculations for
$q=7$. The measure $\mu$ in eq.\eq{eff:potential} was chosen to be
quadratic, $\mu(j_x) = \exp\{ - {1 \over 2 \beta_j} {||j_x||}^2\}$, with
$\beta_j = 0.6$. Effectively we simulated the 4D Abelian Higgs model and for
each configuration of 4D fields we simulated 3D model to get the Mandelstam
strings with the weight $\mu(j_x)$. It was enough for our measurements 20
statistically independent 4D field configurations, and for each of these
configurations we generated 17 configurations of 3D Mandelstam strings. We
imposed the anti-periodic boundary conditions in space.

The effective potential \eq{eff:potential} for positive values of the
monopole field is shown in Figure~\ref{fig:potentials}. In the confinement
phase ($\beta=0.9$) the potential has a Higgs form signalling the monopole
condensation. In the deconfinement phase the potential has minimum at $\Phi
= 0$ which indicates the absence of the monopole condensate.

We conclude that the new operator can be used as a test of the monopole
condensation in the theories with electrically charged matter fields.
The minimum of the potential, corresponding to the value of the monopole
condensate is zero in deconfinement phase (Fig.1(b)) and non zero in
the confinement phase (Fig.1(a)).
\vspace{-1mm}
\section*{ACKNOWLEDGMENTS}
The authors are grateful to F.V. Gubarev for useful
discussions. V.A.B. and M.I.P were partially supported by grants RFBR
00-15-96786, RFBR 01-02-17456, INTAS 00-00111 and CRDF award RP1-2103.
M.I.Ch. is supported by JSPS Fellowship P01023.

\end{document}